\documentclass[twocolumn,twoside,slac_two]{revtex4}
\usepackage{graphicx}
\usepackage{fancyhdr}
\begin{document}

\title{The MAGIC view of PG~1553+113}

\author{E. Prandini}
\affiliation{Padova University \& INFN}
\author{F. Tavecchio}
\affiliation{Osservatorio Astronomico di Brera}

\author{on behalf of the MAGIC Collaboration, S. Buson and S. Larsson on behalf of the {\it Fermi}/LAT Collaboration}
 
\begin{abstract}
We present the results of five years (2005-2009) of MAGIC observations of the BL Lac object PG 1553+113 at very high energies (VHEs). Adding the new data set (2007-2009) to previous observations, this source becomes one of the best long-term followed sources at energies above 100 GeV. In the last three years of data, the flux level above 150 GeV shows a marginal variability. Simultaneous optical data also show only modest variability that seems to be correlated with VHE gamma-ray variability. We also performed a temporal analysis of all available {\it Fermi}/LAT data of PG 1553$+$113 above 1 GeV. Finally, we present a combination of the mean spectrum measured at VHE with archival data available for other wavelengths. The mean Spectral Energy Distribution (SED) can be modeled with a one-zone SSC model, which gives the main physical parameters governing the VHE emission in the blazar jet.

\end{abstract}

\maketitle

\thispagestyle{fancy}

\section{Introduction}
PG~1553$+$113 is a BL Lac object located in the Northern hemisphere. 
It was discovered in 1986 by Green et al. \cite{green86}. 
As for many objects of this class, its redshift is uncertain. Several attempts to determine its distance were done in the past, e.g. \cite{treves07}. Recent determinations suggest $z\,\sim$\,0.4 \cite{danforth10,prandini10}.
At VHE the blazar was detected in 2005  by the H.E.S.S. telescopes system  \cite{aharonian08} and soon after confirmed by MAGIC \cite{albert07}. 

\section{ MAGIC Observations}
\begin{figure*}
\includegraphics[width=135mm]{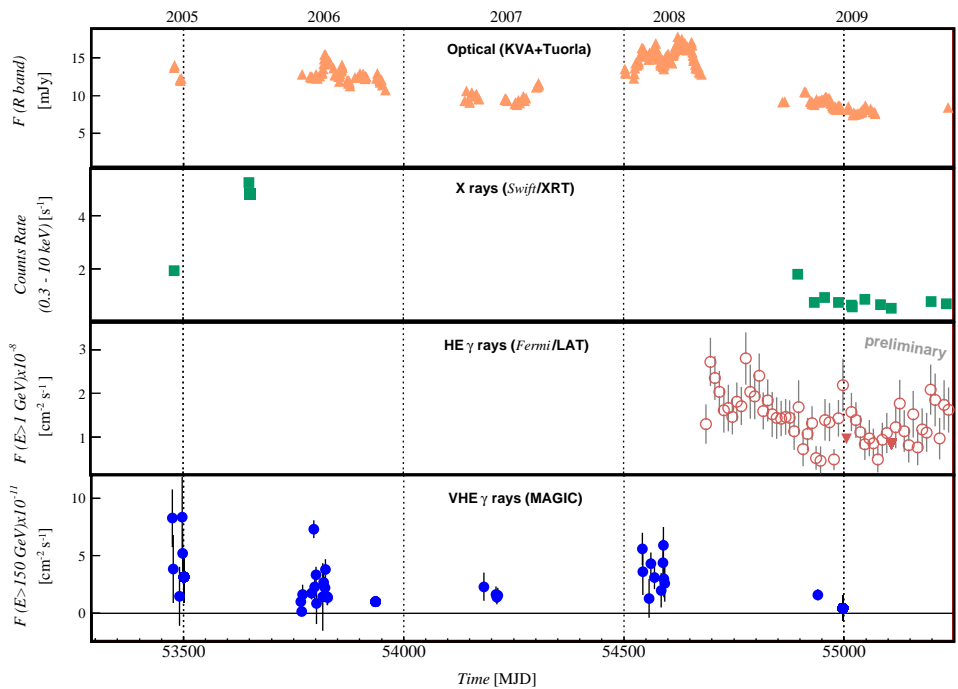}
\caption{ -- Multiwavelength light curve of PG~1553+113 from 2005 to late 2009. From upper to lower panel:  optical flux in the R-band (triangles), X rays counts rate (squares), $\gamma$ rays above 1\,GeV (open circles), and VHE $\gamma$ rays above 150\,GeV measured by MAGIC (filled circles).  The error bars reported have 1\,$\sigma$ significance. Downward triangles, in the third panel, refer to 2\,$\sigma$ upper limits on the source flux above 1\,GeV.}
\label{fig:1553_lc}
\end{figure*}
\begin{table*}[bht]
  \centering
  \begin{tabular}{c|c|c|c|c|c}
  \hline
 \hline
   Year & Good quality data  & Energy Threshold & $F$ ($>$ 150\,GeV) / 10$^{-11}$ &  $F$ ($>$ 150\,GeV) &  $\Gamma$  \\
        & [h]           & [GeV]   & [ cm$^{-2}$ s$^{-1}$]           & [Crab \%]         &   \\
\hline 
\hline
    2005+06 & 19          &  90    & 2.8 $\pm$ 0.5   & 9\% &  4.2 $\pm$ 0.2\\
\hline
    2007 & 11.5           &  80    & 1.40 $\pm$ 0.38 & 4\%  &  4.1 $\pm$ 0.3 \\
 \hline 
    2008 & 8.7 (6.9 for spectrum) &   150  & 3.70 $\pm$ 0.47 & 11\% & 4.3 $\pm$ 0.4 \\
 \hline 
    2009 & 8.5 (6.9 for spectrum) &   160  & 1.63 $\pm$ 0.45 &  5\% &  3.6 $\pm$ 0.5 \\
 \hline 
 \hline  
\end{tabular}  
 \caption{ -- Observation details and spectra of the individual years of observations of  PG~1553+113.}  
  \label{tab:1553_flux}
  \end{table*}
MAGIC is a stereo system composed of two new generation Imaging Atmospheric 
Cherenkov Telescopes located on La Palma, Canary Islands, Spain at 
$\sim$\,2200\,m asl. It observes the VHE $\gamma$-ray sky at energies above
60\,GeV. Data presented here were collected before Autumn 2009, when MAGIC 
was operating with a single telescope. For a detailed description of the 
performances, see \cite{albert08a}.

Since 2005, PG~1553$+$113 was monitored by MAGIC \cite{albert09,aleksic11,aleksic10}. 
Here we present the results of the analysis of 2007/08/09 data.
The data (partly taken in moderate light conditions, i.e. moon light) 
were analyzed  using the standard MAGIC 
analysis chain \cite{albert08a,aliu09}.
Severe quality cuts based on event rate after night sky background
suppression were applied to the sample; 28.7~hours of good quality
data remained after these cuts, out of which 25.3 can be used for the spectral
analysis (Table~\ref{tab:1553_flux}).
The energy threshold is $\sim$\,90 and $\sim$\,80\,GeV for 2006 and 
2007 observations respectively, $\sim$\,150 for 2008, due to 
poor observing conditions, and 160\,GeV for 2009 data, due to moderate moon 
light observation. 
Further details of the data analysis and signal detection of
the sample are discussed in \cite{aleksic11}.

\section{Results}
 \subsection{Integral Flux}

\begin{figure}
\includegraphics[width=70mm]{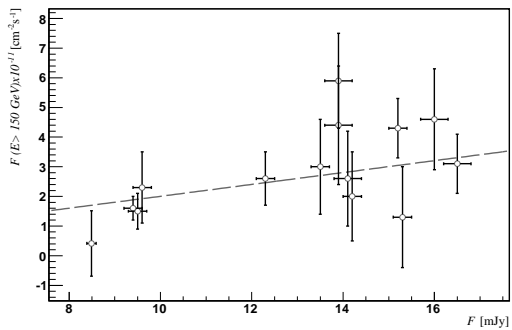}
\caption{ -- Correlation study between PG~1553+113 optical R--band flux and VHE $\gamma$ ray integral flux above 150\,GeV observed from 2006 to 2009.}
\label{fig:1553_correlation}
\end{figure}

The lower panel of Figure~\ref{fig:1553_lc} displays
the VHE integral flux  above 150\,GeV 
of PG~1553+113 measured from 2007 to 2009 by MAGIC with a
variable binning. For comparison, the daily flux levels measured in 
2005 and 2006 are shown, as 
extrapolated from the published data \cite{albert07, albert09}.
The measured flux shows modest variations (4\% to 11\% of the Crab
Nebula flux above 150\,GeV, Table\,\ref{tab:1553_flux}).

The simultaneous optical R-band data are outlined in the first panel. 
These data are collected on a nightly basis by the Tuorla Observatory Blazar 
Monitoring Program\footnote{More information at http://users.utu.fi/kani/1m/} 
\cite{takalo07} using the KVA 35\,cm telescope at La Palma and the Tuorla 1\,meter telescope in Finland. 
A marginal activity is followed by the optical flux, whose variations are  
limited within a factor of four. 
Figure~\ref{fig:1553_correlation} shows the result of a correlation study
between optical and TeV simultaneous observations. 
A linear relation among the two components has a 74\% probability 
($\chi^2$ test),
which suggests a correlation between these two extreme energy bands. 

The public X-ray data, results of an automatic analysis performed by the 
{\it Swift}/XRT Monitoring Program\footnote{http://www.swift.psu.edu/monitoring/}, 
are shown in the second panel of Fig\,\ref{fig:1553_lc}. In contrast to optical and 
VHE bands, the X-ray light curve shows a pronounced variability.

In the third panel, the {\it Fermi}/LAT light curve of PG~1553+113, 
computed in 10-day bins, is displayed. {\it Fermi} data presented 
are restricted to the 1\,GeV\,-\,100\,GeV energy range and were 
collected from MJD 54682 (2008 August 4) to MJD 55200 (2010 January 4) 
in survey mode. The details on the analysis performed can be found in
\cite{aleksic11}. A steady emission above 1\,GeV  has a probability 
smaller than 0.1\% and is ruled out.

\begin{figure*}
\includegraphics[width=75mm]{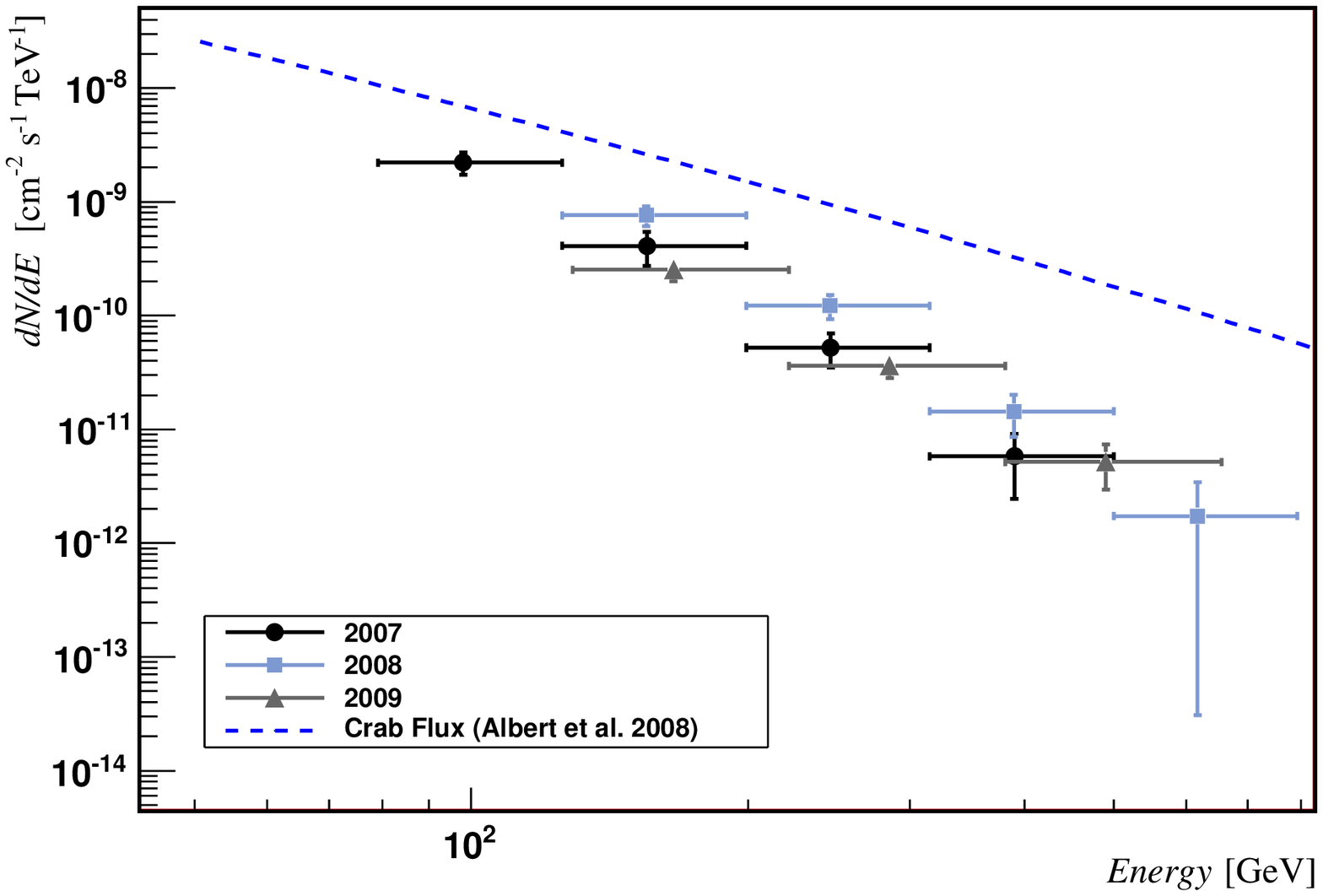}
\includegraphics[width=75mm]{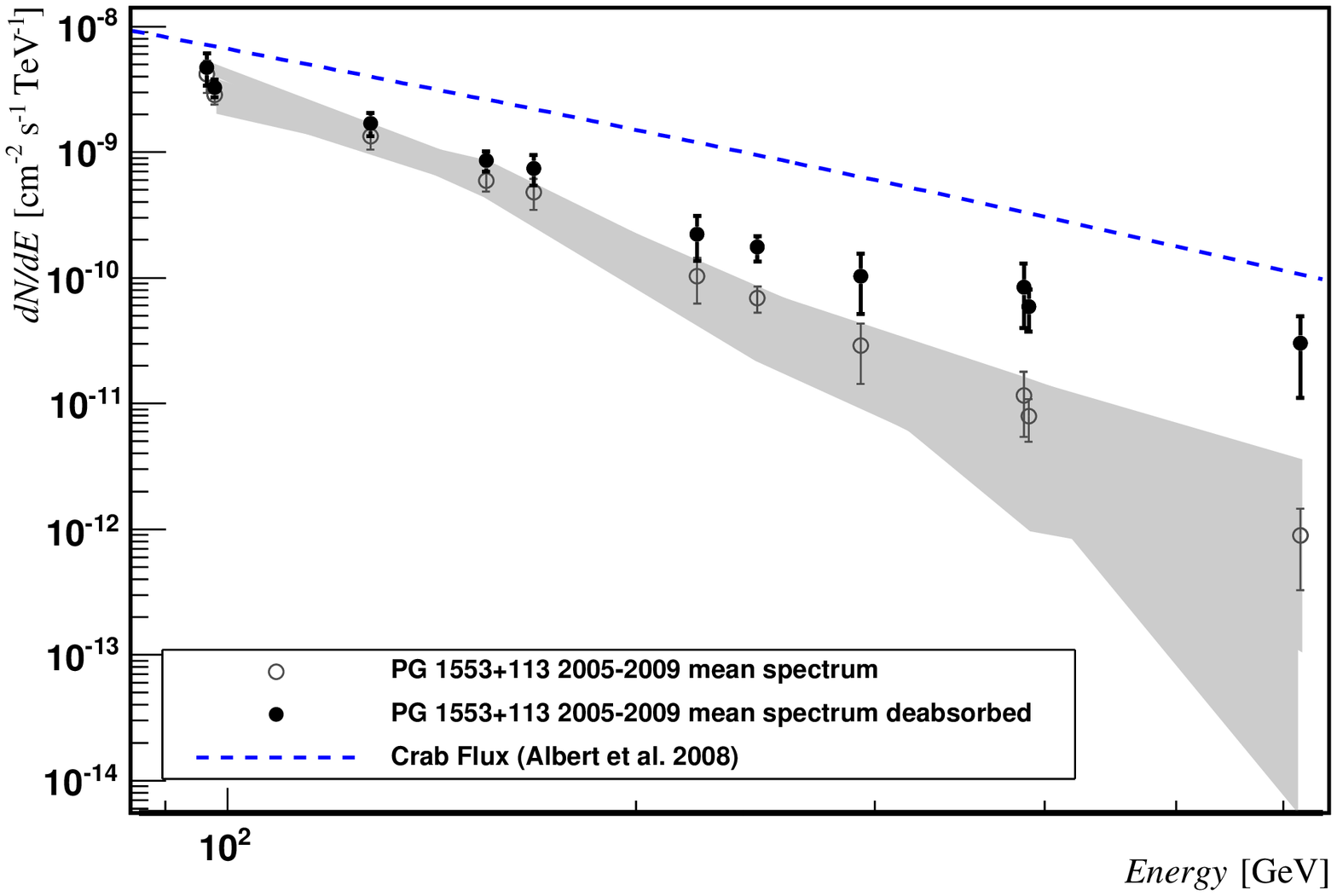}
\caption{ -- Differential energy spectra from PG~1553+113. Left Figure: comparison between 2007, 2008 and 2009 spectra. Right Figure: superimposition of 2005-2006 spectrum, from \cite{albert07}, to 2007-2009 mean spectrum and corresponding deabsorption for {\it z} = 0.4 using the  EBL model of \cite{dominguez11}.
 In both figures, the fit of the Crab Nebula spectrum measured by  MAGIC \cite{albert08a} is superimposed for comparison.}
\label{fig:1553_spectra}
\end{figure*}

\subsection{Differential Flux}
The differential energy spectra observed from PG~1553+113 by MAGIC each
year from 2007 to 2009 are shown in the left plot of Figure~\ref{fig:1553_spectra}.
As for other blazars, each spectrum can be well fitted with a power law function.
The resulting indices are listed in the last
column of Table~\ref{tab:1553_flux}.
The systematic uncertainty is estimated to be $35\%$ in the
flux level and 0.2 in the power index \cite{albert08a}. 
Except for 2009 sample, whose
significance is rather low and
corresponding errors noticeably large, the indices
describing the spectra  are compatible. 
This indicates that the shape of the emitted spectrum does not
change, even if the total flux shows hints of (small amplitude) variability.

The right plot of Figure~\ref{fig:1553_spectra} 
shows the combined differential spectrum of PG~1553+113 from 2007 to 2009,
superimposed to the 2005-2006 spectrum
measured by MAGIC ($\Gamma$~=~4.21~$\pm$~0.25, \cite{albert07}).
The gray band represents the systematic effect on the combined spectrum
result of different unfolding methods.
In order to estimate the intrinsic spectrum emitted by the source,
the effect of absorption of  VHE photons  in the interaction with the 
extragalactic background light (EBL) should be considered. 
Assuming the background model proposed in \cite{dominguez11} and a redshift {\it z}~=~0.40, 
we obtain an intrinsic spectrum compatible with a power law of index 3.09~$\pm$~0.20, 
as drawn in the right plot of Figure~\ref{fig:1553_spectra}.

\begin{figure*}
  \centering 
  \vskip -1.5 truecm
  \includegraphics[width=140mm]{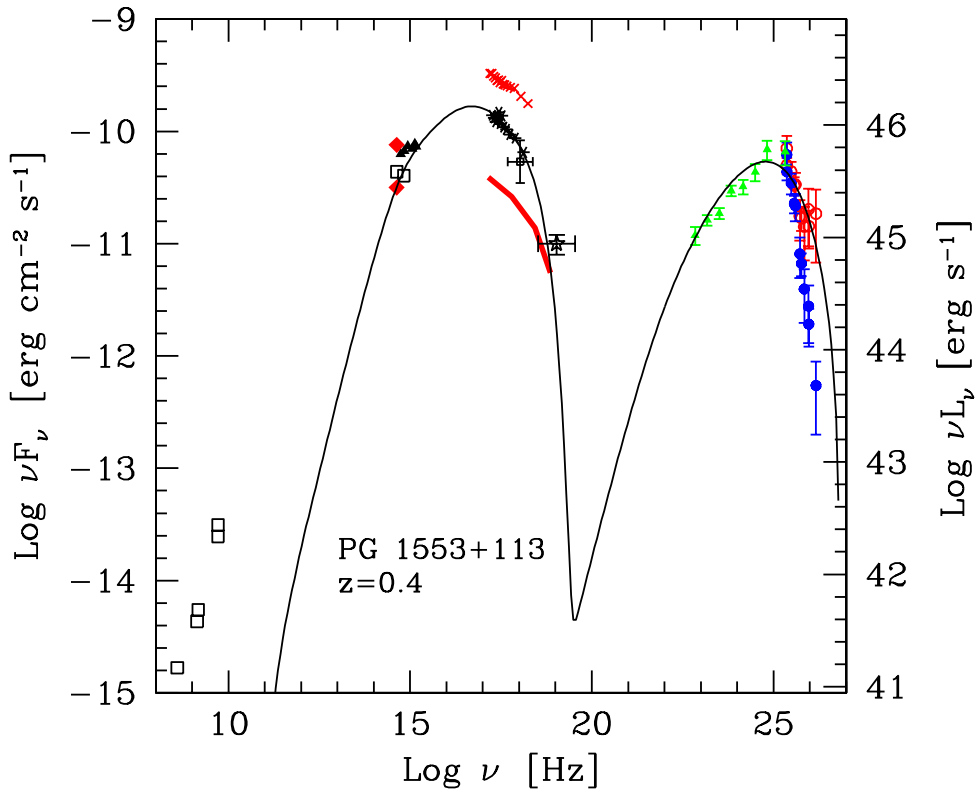}
  \vskip -1.7 truecm
  \caption{ -- SED of PG 1553+113. Open black squares displaying radio-optical  data are from NED. In the optical band, we also show (red diamonds) the KVA  minimum and maximum flux measured in the period covered by MAGIC 2005-2009 observations together with optical-UV fluxes from {\it Swift}/UVOT (filled black triangles, from \cite{tavecchio10}). For the X-ray data, two {\it Swift}/XRT spectra taken in 2005 (high flux state, red crosses, and intermediate state, black asterisks, from \cite{tavecchio10}) are given, and a {\it Suzaku} spectrum taken in 2006 (continuous red line, from \cite{reimer08}). In addition, the average 14-150 keV flux measured by {\it Swift}/BAT  during the first 39 months of survey \cite{cusumano10} is shown (black star),  and the average {\it RXTE}/ASM flux between   March 1 and May 31, 2008 (small black square), from quick--look results  provided by the {\it RXTE}/ASM team.  The green triangles correspond to the LAT spectrum averaged over $\sim$\,200 days (2008 August-2009  February) from \cite{abdo10b}.   For MAGIC, we report the  2005-2006 and 2007-2009   observed spectra (filled circles) and the same spectra corrected for   the absorption by the EBL (red open circles).}\label{fig:1553_SED}
\end{figure*}

\begin{table*}[th]
\centering
\begin{tabular}{lcccccccccccc}
\hline
\hline
$\gamma _{\rm min}$ & $\gamma _{\rm b}$ & $\gamma _{\rm max}$ & $n_1$ & $n_2$ &$B$ & $K$ &$R$ & $\delta $ &$P_{\rm e}$ & $P_{B}$ & $P_{\rm p}$ & $L_{\rm r}$ \\
$[10^3$] & [$ 10^4$] &[$ 10^5$]  &  & &[G] & [$ 10^3$ cm$^{-3}]$  & $[10^{16}$ cm] & &  [$10^{44}$ erg/s] & [$10^{44}$ erg/s]& [$10^{44}$ erg/s] & [$10^{44}$ erg/s]\\
\hline
$2.5$ & $3.2$ & $2.2$ & $2.0$ & $4.0$ & $0.5$ & $5.35$ & $1$ & $35$  & 2.2 & 1.5& 0.34 &  6.3\\
\hline
\hline
\end{tabular}
\vskip 0.4 true cm
\caption{ -- Input model parameters for the model shown in Fig~\ref{fig:1553_SED}. See text for details.}
\label{tab:1553_SED}
\end{table*}

\subsection{Modeling the SED}
Figure~\ref{fig:1553_SED} shows the SED of PG~1553+113 
obtained using historical data and the MAGIC 
spectra described above. 

The mean overall SED can be fitted with a simple one-zone SSC model 
\cite{maraschi03}. The corresponding model
 parameters are  the minimum, break and maximum electron Lorentz factors and
 the low and high energy slope of the electron energy distribution, 
the magnetic field intensity, the electron density, the radius of the 
emitting region and its Doppler factor, listed in Tab \ref{tab:1553_SED}. 
We also give the derived power carried by electrons, magnetic field, protons 
(assuming one cold proton per emitting relativistic electron) and the total radiative luminosity.

\bigskip 
\begin{acknowledgments}
 We would like to thank the Instituto de Astrof\'{\i}sica de
Canarias for the excellent working conditions at the
Observatorio del Roque de los Muchachos in La Palma.
The support of the German BMBF and MPG, the Italian INFN, 
the Swiss National Fund SNF, and the Spanish MICINN is 
gratefully acknowledged. This work was also supported by 
the Marie Curie program, by the CPAN CSD2007-00042 and MultiDark
CSD2009-00064 projects of the Spanish Consolider-Ingenio 2010
programme, by grant DO02-353 of the Bulgaria
n NSF, by grant 127740 of 
the Academy of Finland, by the YIP of the Helmholtz Gemeinschaft, 
by the DFG Cluster of Excellence ``Origin and Structure of the 
Universe'', by the DFG Collaborative Research Centers SFB823/C4 and SFB876/C3,
and by the Polish MNiSzW grant 745/N-HESS-MAGIC/2010/0.

The {\it Fermi} LAT Collaboration acknowledges support from a number of agencies and institutes for both development and the operation of the LAT as well as scientific data analysis. These include NASA and DOE in the United States, CEA/Irfu and IN2P3/CNRS in France, ASI and INFN in Italy, MEXT, KEK, and JAXA in Japan, and the K.~A.~Wallenberg Foundation, the Swedish Research Council and the National Space Board in Sweden. Additional support from INAF in Italy and CNES in France for science analysis during the operations phase is also gratefully acknowledged.
\end{acknowledgments}

\bigskip 

\end{document}